\documentclass[sigplan,screen,sigconf]{acmart}

\usepackage{ulem}
\usepackage{url}

\setcopyright{rightsretained}
\acmDOI{}
\acmISBN{}
\acmConference[LATTE '21]{1st Workshop on Languages, Tools, and Techniques for Accelerator Design}{April 15, 2021}{Virtual, Earth}

\title[High-Level Synthesis of Security Properties 
via Software-Level Abstractions]{High-Level Synthesis of Security Properties\\
via Software-Level Abstractions}

\author[Christian Pilato and Francesco Regazzoni]{Christian Pilato$^{1}$ and Francesco Regazzoni$^{2,3}$}
\affiliation{
    $^{1}$Politecnico di Milano --- DEIB, Milan, Italy, christian.pilato@polimi.it\\
    $^{2}$University of Amsterdam, Amsterdam, The Netherlands, f.regazzoni@uva.nl\\
    $^{3}$Università della Svizzera italiana, Lugano, Switzerland, francesco.regazzoni@usi.ch
    \country{~}
}

\begin{document}

\begin{abstract}
High-level synthesis (HLS) is a key component for the hardware acceleration of applications, especially thanks to the diffusion of reconfigurable devices in many domains, from data centers to edge devices. HLS reduces development times by allowing designers to raise the abstraction level and use automated methods for hardware generation. Since security concerns are becoming more and more relevant for data-intensive applications, we investigate how to abstract security properties and use HLS for their integration with the accelerator functionality. We use the case of dynamic information flow tracking, showing how classic software-level abstractions can be efficiently used to hide implementation details to the designers.
\end{abstract}

\maketitle

\section{Introduction}

The future of computing systems will be necessary data-driven. Collecting and processing large amounts of data will unleash unprecedented knowledge discovery that can improve everyday's life. However, these applications demand not only novel and heterogeneous architectures to delivery energy-efficient high performance but also effective methods to avoid unauthorized operations on the data~\cite{pilato2021everest}.
On one side, \textit{high-level synthesis} (HLS) is a key enabler for heterogeneous architectures. Abstracting the functionality of a component to the software level and applying automated methods for hardware generation, HLS allows non-expert designers to create more components, specialize their architectures, and reduce design costs. We expect more and more HLS-generated components to be integrated in future systems.
On the other hand, dealing with valuable data attracts \textit{malicious actors} that can steal (or alter) sensitive information or use the existing data flow to compromise the system. For example, buffer overflow is a  technique to exploit software vulnerabilities to gain control of an application and potentially steal sensitive data. While hardware-assisted security protections are more efficient, their implementation requires to modify the components or the design flow. Integrating data and intellectual property (IP) protection into HLS is interesting~\cite{Pilato2018} but previous attempts require extensive tool modifications~\cite{Pilato2019,Pilato_dac_2018,8715199,8587747} or are limited to specific security properties~\cite{date21badier}.

In this line of research, we are exploring which security protections can be specified at the software level and synthesized transparently during HLS.

\section{Compiler Infrastructure for HLS}

\begin{figure}[!ht]
\includegraphics[height=2.2cm]{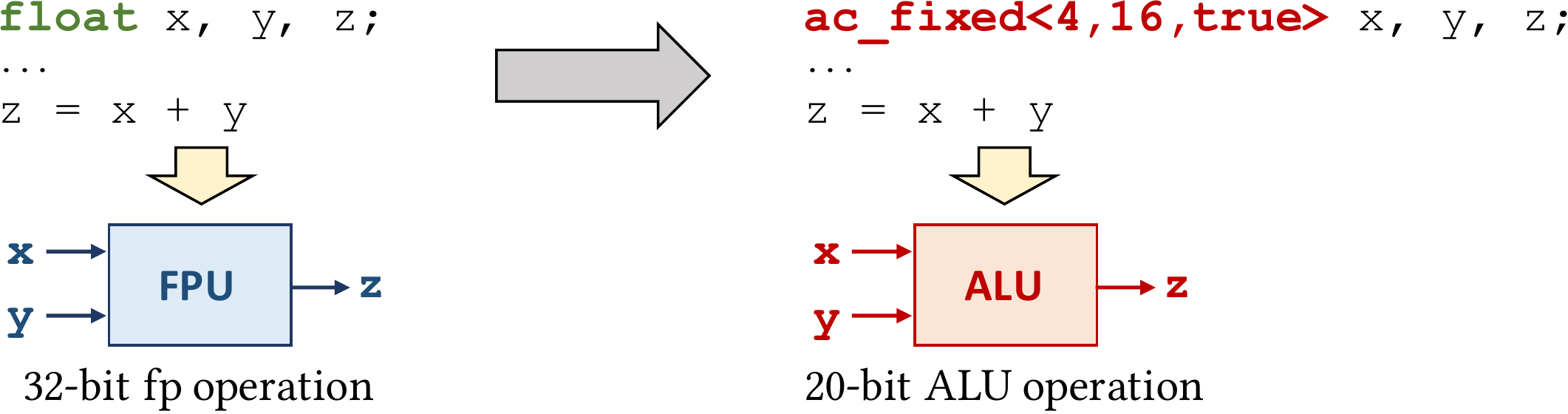}
\vspace{-8pt}\caption{Example of custom datatypes for HLS.}\label{fig:datatypes}
\end{figure}

Modern HLS tools leverage state-of-the-art compilers like GCC or LLVM as a frontend to software specifications~\cite{10.1109/TCAD.2015.2513673}. Such compilers extract a language-agnostic representation with the essential semantics to synthesize in hardware. They are also used to apply code transformations (e.g., loop optimizations and constant propagation), create a more hardware-friendly description (e.g., code lowering and bit-width optimization), and extract more hardware parallelism (e.g., inlining and memory optimizations). In the following steps, the HLS engine performs temporal and spatial assignment of the operations to derive the corresponding microarchitecture.
Software abstractions are widely used to create compact but flexible representations and to hide details to the programmers. Among those, \textbf{synthesizable software libraries} and \textbf{operator overloading} are common also in HLS. Software libraries can abstract common hardware functions like recurrent functions~\cite{8412613}, memory data transfers, and communication protocols. For example, \texttt{hlslib}~\cite{hlslib} provides libraries to support designers in common optimization steps, like interface and communication synthesis.
Operator overloading can associate different implementations to the same operators based on their arguments. For example, Mentor offers \texttt{ac\_types} that are bit-accurate datatypes for custom precision~\cite{ac_types}. \autoref{fig:datatypes} shows an example for converting floating-point to fixed-point operations.

\begin{figure*}[t]
\includegraphics[width=\textwidth]{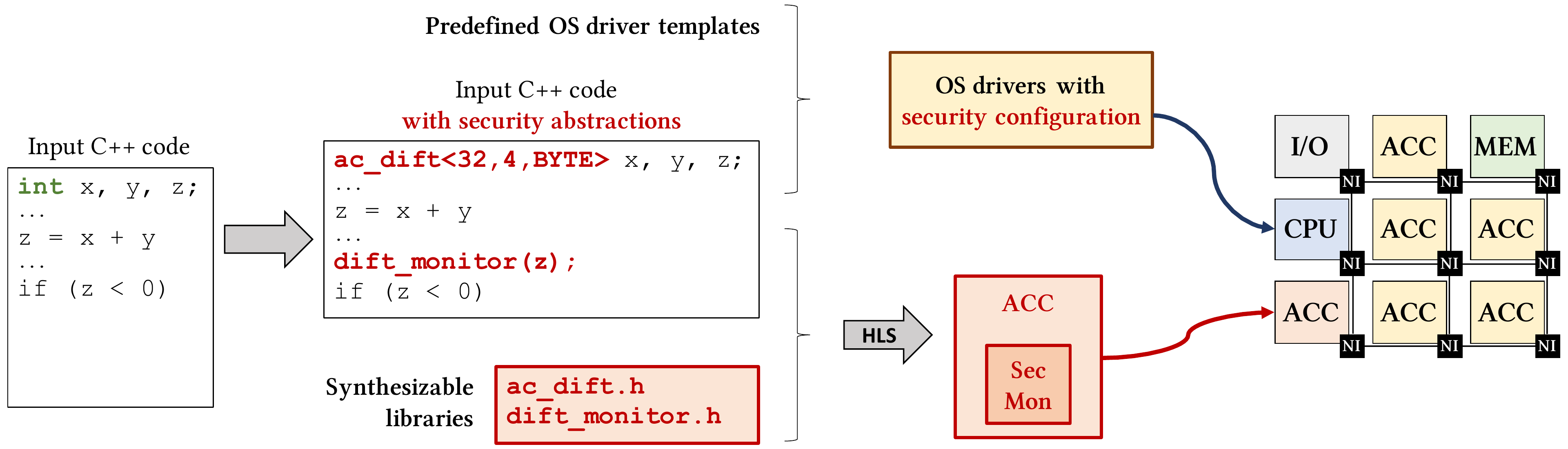}
\caption{Software-level abstraction to implement synthesizable dynamic information flow tracking.}\label{fig:abstractions}
\end{figure*}

While these solutions are commonly used for optimizing the microarchitecture of hardware accelerators, their adoption for integrating security features is still at the initial stages. Indeed, integrating specialized security components and automatically propagating security properties can be achieved with this approach. However, HLS needs to be carefully tuned to optimize the logic while, at the same time, avoid introducing hardware-level vulnerabilities, like power side-channels~\cite{8887278}.
 
\section{Propagation of Security Properties}

While HLS is good at optimizing classic non-functional requirements (e.g., area, power, and delay), the propagation of security properties and the integration of security protections need more careful investigation. We use the case of \textbf{dynamic information flow tracking} (DIFT) as a paradigmatic example. DIFT associates a \textit{tag} to selected data of an application to monitor their influence on the program execution and, ultimately, detect security hazards~\cite{10.1145/1037187.1024404}. The integration of specific security policies can help contain the effects of these hazards.

Implementing information flow tracking in hardware is complex and expensive. Researchers proposed several solutions to trade-off accuracy of the taint analysis and hardware cost~\cite{kastner_survey_2020}.
Coarse-grained approaches apply DIFT to the ``boundaries'' of the components~\cite{8412617,6658991}, while fine-grained taint propagation leads to high area overhead~\cite{5948366} or requires tool modifications to automatically integrate the additional logic~\cite{Pilato2019}. Furthermore, implementing security policies require to generate and integrate proper hardware monitors and assertion-like logic~\cite{7503118}.
Designers can use HLS to automatically generate such additional logic,  explore the design space of these solutions, and identify the best design point for the target application. HLS can also automatically handle the generation of the monitor components. However, abstracting both hardware and security details is critical for non-expert hardware designers.

\section{Software-Level Abstractions for Accelerator-Level DIFT}

Since all DIFT elements (tags and propagation rules) are related to additional functionalities of the accelerators, we argue they can be embedded in the initial code with minimal changes and, leveraging a combination of synthesizable libraries and HLS, we can automatically generate DIFT-enhanced accelerators. \autoref{fig:abstractions} shows how software-level abstractions can be used to embed DIFT in HLS-generated accelerators. 

The input C++ code is modified by the designer to annotate DIFT variables. We use the custom datatype \texttt{ac\_dift}, where the designer specifies the precision of the variable (like, for example, in \texttt{ac\_int}), the size of the taint tag, and the type of propagation rules to be used. An additional value provides the default taint information. The HLS tool uses the \textbf{custom datatype definition} to include the additional taint variables and the \textbf{operator overloading} to synthesize the logic that combines the operator variables and their taint values. In this way, it can automatically compute the operation result along with the associated tag according to the given propagation rule.

The designer also modifies the input code to include \textit{security checkpoints} with specific function calls. The security policies are implemented as \textbf{synthesizable libraries}. For example, the function \texttt{dift\_monitor(x)} requests to check the tag of  variable \texttt{x}. After applying HLS on the function augmented with the security checkpoint, the accelerator will include a \textbf{security monitor}, i.e. a submodule that receives the taint tags and produces control signals based on the tag values and the given security policy. The designer can customize the security policy by specializing the function \texttt{dift\_monitor}. In case of security hazards, the monitor produces a \textit{security exception} that is trapped and managed by the system. The exception can trigger, for example, a special \textit{interrupt request} that activates system-level protections like component isolation. Predefined Operating System (OS) drivers are customized with the proper configuration of I/O registers to exchange tag information with software~\cite{10.1145/3400302.3415753}.

This approach has several potential advantages that are worth to be explored. First, it provides a complete infrastructure that provides implementation support for non-experts. Second, it allows the designers to check DIFT and security policies at a higher level of abstraction, along with the rest of the software code. Third, the DIFT functions inside operator overloading and the monitor libraries are synthesized (and co-optimized) along with the rest of the accelerator's logic. The HLS engine could introduce extra cycles to optimize the schedule and minimize resource utilization, without a perfect \textit{data flow consistency} between baseline and DIFT microarchitectures~\cite{Pilato2019}. However, these optimizations would not affect the DIFT results.

\section{Conclusion}

We discuss code-level extensions to specify security protections that can be later automatically synthesized with HLS. For this, we analyze the case of dynamic information flow tracking and how software-level abstractions can support the designers. This activity opens up an interesting research question: \textit{\uline{Which security protections can be effectively abstracted and synthesized with HLS without compromising their security?}}

\section*{Acknoledgements}

This project is partially funded by the EU Horizon 2020 Programme under grant agreement No 957269 (EVEREST).

\newpage

\balance
\bibliography{main}

\end{document}